\def \SAIT #1 #2 {{\em Mem.\ Soc.\ Astron.\ It.\/} {\bf #1}, #2}
\def \MESS #1 #2 {{\em The Messenger\/} {\bf #1}, #2}
\def \ASTRNACH #1 #2 {{\em Astron. Nach.\/} {\bf #1}, #2}
\def \AAP #1 #2 {{\em Astron. Astrophys.\/} {\bf #1}, #2}
\def \AAL #1 #2 {{\em Astron. Astrophys. Lett.\/} {\bf #1}, L#2}
\def \AAR #1 #2 {{\em Astron. Astrophys. Rev.\/} {\bf #1}, #2}
\def \AAS #1 #2 {{\em Astron. Astrophys. Suppl. Ser.\/} {\bf #1}, #2}
\def \AJ #1 #2 {{\em Astron. J.\/} {\bf #1}, #2}
\def \ANNREV #1 #2 {{\em Ann. Rev. Astron. Astrophys.\/} {\bf #1}, #2}
\def \APJ #1 #2 {{\em Astrophys. J.\/} {\bf #1}, #2}
\def \APJL #1 #2 {{\em Astrophys. J. Lett.\/} {\bf #1}, L#2}
\def \APJS #1 #2 {{\em Astrophys. J. Suppl.\/} {\bf #1}, #2}
\def \APSS #1 #2 {{\em Astrophys. Space Sci.\/} {\bf #1}, #2}
\def \ASR #1 #2 {{\em Adv. Space Res.\/} {\bf #1}, #2}
\def \BAIC #1 #2 {{\em Bull. Astron. Inst. Czechosl.\/} {\bf #1}, #2}
\def \JSQRT #1 #2 {{\em J. Quant. Spectrosc. Radiat. Transfer\/} {\bf #1}, #2}
\def \MN #1 #2 {{\em Mon. Not. R. Astr. Soc.\/} {\bf #1}, #2}
\def \MEM #1 #2 {{\em Mem. R. Astr. Soc.\/} {\bf #1}, #2}
\def \PLR #1 #2 {{\em Phys. Lett. Rev.\/} {\bf #1}, #2}
\def \PASJ #1 #2 {{\em Publ. Astron. Soc. Japan\/} {\bf #1}, #2}
\def \PASP #1 #2 {{\em Publ. Astr. Soc. Pacific\/} {\bf #1}, #2}
\def \NAT #1 #2 {{\em Nature\/} {\bf #1}, #2}
\documentstyle{memsait}
\input epsf.sty
\begin{opening}
\title{The \lq{Tip}\rq\ of the Red Giant Branch as a distance indicator: theoretical
calibration and the value of $H_{0}$}
\author{Santi Cassisi$^{1,2,3}$, Maurizio Salaris$^3$}
\institute{$^1$Osservatorio Astronomico di Collurania, I-64100, Teramo, Italy\\
$^2$Universit\'a degli studi de L'Aquila, Dipartimento di Fisica, I-67100, L'Aquila, Italy\\
$^3$Max-Planck-Institut f\"ur Astrophysik, D-85740,
Garching, Germany }

\date{} % DO NOT INSERT ANY DATE HERE !!!
\end{opening}

\begin{document}

%\oddpagefooter{\sf Mem. S.A.It., Vol. ??, ??}{}{\thepage}
%\evenpagefooter{\thepage}{}{\sf Mem. S.A.It., Vol. ??, ??}
\oddpagefooter{}{}{} % LEAVE AS IT IS !
\evenpagefooter{}{}{} % LEAVE AS IT IS !
%\ 
%\bigskip
%
\begin{abstract}

Updated theoretical relations for the run of the 
bolometric and I magnitude of the Tip of the Red Giant Branch (TRGB)
with respect to the metallicity of the parent stellar population are provided.
An analogous relation for the V magnitude of the Zero Age Horizontal Branch
(ZAHB) at the RR Lyrae instability strip is also provided.

A comparison has been performed among our ZAHB and TRGB distances,
the Cepheid distance scale by Madore \& Freedman (1991) and the 
HIPPARCOS distances set by local subdwarfs with accurate parallax determinations.
The ZAHB, TRGB and HIPPARCOS distances are in satisfactory agreement,
whereas the comparison between TRGB and Cepheid distances
discloses a systematic discrepancy of the order of 0.12 mag, the TRGB distances 
being systematically higher. This result supports the case for a revision of the 
zero point of the Cepheid distance scale.

The application of our TRGB distance scale to NGC3379 provides a distance
to the Leo I group that is about 8\% higher than the one obtained by
Sakai et al. (1997) adopting the TRGB brightness calibration by 
Da Costa \& Armandroff (1990). Our distance to the Leo I group, coupled 
with the relative distance Coma cluster-Leo I determined 
differentially by means of secondary distance indicators, provides a determination
of $H_{0}$ at the Coma cluster: $H_{0}$=64$_{-9}^{+10}$ Km $s^{-1} Mpc^{-1}$.

\end{abstract}

\section{Introduction.}

The Tip of the 
Red Giant Branch (TRGB) method has been recently used for estimating the distances
to several nearby galaxies (see, e.g, the list in Salaris \& Cassisi 1997);
Lee, Freedman \& Madore (1993 - hereinafter LFM93)
and Madore \& Freedman (1995) assessed the reliability and
intrinsic accuracy of this method, and demonstrated
that the TRGB can be successfully used for determining
distances accurate within 0.2 mag for galaxies out to 3 Mpc by using ground-based
telescopes, and out to 12 - 13 Mpc by using the {\sl HST}.

The underlying physical mechanism which allows to use the TRGB as a standard
candle is the following: the TRGB marks the Helium ignition inside the
degenerate core of low-mass stars, and its brightness depends
on the He core mass, which
is remarkably constant for ages larger than a few billions of years.

A fundamental ingredient for using the TRGB as a distance indicator is
the calibration of its bolometric magnitude ($M_{bol}^{TRGB}$) as a function of 
[M/H]. The relation generally used until now
is the semiempirical one by Da Costa \& Armandroff (1990 - hereinafter
DA90), based on the 
observational database by Frogel, Persson \& Cohen (1983 - hereinafter
FPC83) of bolometric magnitudes for RGB stars in a sample
of galactic globular clusters (hereinafter GC).
The observed $m_{bol}$ of the most luminous red giants 
are converted into 
absolute magnitudes adopting distance moduli for the parent GC
obtained by using the RR Lyrae distance scale by Lee, Demarque \& Zinn
(1990), and these absolute magnitudes fix the zero point of the
TRGB distances. 
As discussed in Salaris \& Cassisi (1997) this procedure provides a
zero point too faint (independently of the accuracy of the Lee,
Demarque \& Zinn RR Lyrae distance scale); the reason is the small
sample of stars observed by FPC83. Taking into account the
evolutionary times along the RGB and the observed sample of stars,
it is possible to compute statistically the probability that the most 
luminous observed star is actually at the TRGB; the result is that
this probability is very small,
indicating clearly that the TRGB brightness is systematically 
underestimated.

In this paper we will present a purely theoretical calibration of
the TRGB brightness as a function of [M/H] as obtained
from updated evolutionary computations, and we will compare
the distance scale set by our TRGB luminosities with 
RR Lyrae, HIPPARCOS and Cepheid distances. Once assessed the reliability
of our calibration, we will apply the TRGB method for the
determination of the Hubble constant.

\section{The TRGB distance scale}
\subsection{Theoretical stellar models}

We have determined the TRGB luminosities 
for stellar populations with age t=15 Gyr 
(but, as discussed
before and in Salaris \& Cassisi 1997, the precise value of t does not influence
the TRGB luminosities for ages larger than a few Gyr) and metallicity
$-2.35\le[M/H]\le-0.28$ (the He abundance is set to Y=0.23 with the
exception of the case with
[M/H]=-0.28, where we adopted Y=0.255), by computing evolutionary tracks of
low-mass stars without chemical elements diffusion.
As far as it concerns the physical inputs adopted in computing the stellar models,
the interested reader is referred to Salaris \& Cassisi (1997).
Assuming for the Sun $M_{Bol,\odot}=4.75$ mag, we obtain the following
relation:
\begin{equation}
{M^{TRGB}_{Bol}}=-3.949 - 0.178\cdot[M/H] + 0.008\cdot{[M/H]}^2 
\label{mboltip}
\end{equation}
This relation takes also
automatically into account the enhancement of the $\alpha$ elements
observed in galactic field halo and GC stars
when considering the global metallicity [M/H] (see Salaris, Chieffi \&
Straniero 1993).
It is moreover possible to correct this relation for different He contents
around the values we have adopted, taking into account that on average
${\partial{M_{Bol}^{tip}}}\over{\partial{Y}}$ is $\approx1.0$ in the
metallicity range covered by Equation 1.

In order to estimate the internal accuracy of the current theoretical scenario,
in Figure 1 we have compared our Equation \ref{mboltip}
with similar relations derived by independent updated stellar models.
We compared our results with the ones by
Cassisi et al. (1997 - their 'step8', with and without He and heavy
elements diffusion), Caloi et al. (1997 - no diffusion) and Straniero et al. 
(1997 - no diffusion). 
In the same figure the relation provided by DA90 is shown after 
correcting for the slightly different $M_{Bol,\odot}$ adopted by the quoted authors.

It is important to note that 
the agreement between the different recent evolutionary results is
quite good. All the theoretical relations lie within $\approx \pm$0.05
mag with respect to Equation \ref{mboltip}.
Moreover, the change of the TRGB brightness due to the inclusion of
atomic diffusion - adopting the same physical
inputs as in standard models - results to be quite negligible (see the results
corresponding to the Cassisi et al. 1997 models).
Finally, it exists a systematic difference by about 0.15 mag in the zero point between our
relation and the relation provided by DA90
(while the slopes are in good agreement), our TRGB brightnesses being higher.

\begin{figure}
\epsfxsize=6.7cm % fix the x-dimension and scales y-dim. to x-dim.
\hspace{3.8cm}\epsfbox{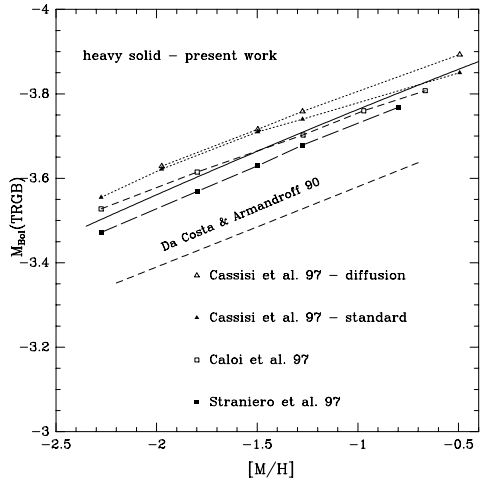}  
\caption[h]{Comparison of updated theoretical relations between TRGB
bolometric magnitude and [M/H]. The calibration provided by DA90 is
also plotted. In all cases, $M_{Bol,\odot}= 4.75$ mag has been adopted.}
\end{figure}

\subsection{The Bolometric Correction scale}

Before going on, we wish to briefly review the iterative procedure suggested by
LMF93 in order to derive distances by means of the TRGB method from observations in the 
VI Johnson-Cousins bands of resolved galaxies. The first step consists
in fixing a preliminar distance modulus;
with this fixed distance modulus one determines the metallicity by
measuring the dereddened $(V-I)$ color at $M_{I}=-3.5$ mag
($(V-I)_{0,-3.5}$) and using a
relation between this color and the metallicity of the parent stellar
population (see section 3.1). As a second step, with this estimate of [M/H] and the
observed I magnitude of the TRGB (corrected for the interstellar extinction), 
one redetermines the distance modulus by adopting a
relation for both the TRGB bolometric magnitude as a function of metallicity 
and the bolometric correction to the I magnitude ($BC_{I}$). At this point,
one iterates the procedure until convergency is achieved.
Due to the weak dependence of $M_{I}^{tip}$ on the metallicity,
convergence is generally achieved after one iteration.

For applying this procedure, it is therefore necessary to have
a relation providing the bolometric correction to the I (Cousins) band.
Following LFM93, an empirical $BC_{I}-(V-I)_{0}$ relation for RGB stars has been
taken from DA90 ($(V-I)_{0}$ is the dereddened color of the
considered RGB stars): $BC_{I}= 0.881- 0.243\cdot(V-I)_0$,
independent of the metallicity.
This empirical relation was derived by comparing the I
magnitudes given in DA90 with the bolometric magnitudes given by
FPC83 for a sample of RGB stars in 8 GC with
different metallicities. By examining Figure 14 in DA90, it appears
clearly that in the range of $(V-I)_{0}$ values typical of the stars
considered by the authors ($(V-I)_{0}$ colors between 1.0 and 1.6) and
of the TRGB stars in the sample of galaxies studied in section 3.2
($(V-I)_{0}$ colors between 1.3 and 2.0),
there is a dispersion of the order of 0.10 mag around the least square
fit that they give. Moreover, the relation for the reddest stars is
based only on a very small number of observational points.
We have therefore used two other independent sets of $BC_{I}$
for better assessing the uncertainty in the TRGB distances due to
the bolometric correction scale.

By using   
the theoretical bolometric corrections by Castelli, Gratton \& Kurucz
(1997a, 1997b - hereinafter K97) and the semiempirical ones by
Green (1988 - hereinafter Yale transformations), we have obtained the
following relations:

\begin{equation}
M_{I}^{TRGB,K97}=-3.953 + 0.437\cdot[M/H] + 0.147\cdot[M/H]^2
\label{mik97}
\end{equation}

\begin{equation}
M_{I}^{TRGB,Yale}=-4.156 + 0.157\cdot[M/H] + 0.070\cdot[M/H]^2 
\label{miyale}
\end{equation}

All these three different sets of $BC_{I}$ will be used in the next sections
for deriving TRGB distances to a sample of resolved galaxies.

\section{Comparison among TRGB, RR Lyrae, HIPPARCOS and Cepheid distance scales}

\subsection{Globular Clusters}

In the case of galactic GC it is possible to compare the 
RR Lyrae distance scale with the one derived from Equation \ref{mboltip}.
For determining the RR Lyrae distance scale
we have here adopted the Zero Age Horizontal Branch 
(hereinafter ZAHB) models from Cassisi \& Salaris (1997), homogeneus
with the models adopted for deriving the TRGB luminosities.
Our ZAHB models have been transformed into the observational plane
by using both the Yale and K97 transformations.
The relations between the ZAHB V magnitude (taken at
$\log{T_{eff}}=3.85$, that corresponds approximately to the average
temperature of the RR Lyrae instability strip)
and [M/H] obtained from these two sets of transformations agree within
0.03 mag. In the following we adopt for fixing the RR Lyrae distance scale
the relation (valid for $-2.35\le[M/H]\le-0.57$):

\begin{equation}
M^{zahb}_{V}= 0.921 + 0.329\cdot[M/H] + 0.045\cdot[M/H]^2
\label{rryale}
\end{equation}
  
The TRGB bolometric magnitudes published by FPC83 
for a sample of GC with accurate spectroscopic determinations of [M/H]
(see Salaris \& Cassisi 1996) have been corrected for our ZAHB distance scale
(see Cassisi \& Salaris 1997 and Salaris \& Cassisi 1997 for details
about the procedure followed for determining the observational ZAHB
level at the instability strip), so that the comparison
displayed in Figure 2 between the observational $M_{Bol}^{tip}$ values
and our Equation  \ref{mboltip} is a comparison between our ZAHB and TRGB distances.
The vertical error bar ($\pm0.1$ mag) for the observational points 
represents an average error on the distance modulus obtained from
Equation \ref{rryale} while
the error on the spectroscopic determination of [M/H] is typically of the order of
0.15 dex.
\begin{figure}
\epsfxsize=6cm % fix the x-dimension and scales y-dim. to x-dim.
\hspace{4.5cm}\epsfbox{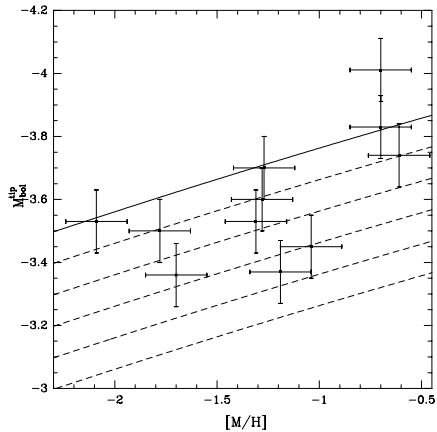} 
\caption[h]{The absolute bolometric magnitude of the brightest observed red giant
as a function of the global metallicity,
for the sample of clusters selected from the FPC83 database. 
The solid line shows the theoretical expectation for the bolometric
magnitude of the TRGB; the dashed lines represent the same theoretical relation but
shifted in steps of 0.1 mag.}
\end{figure}
Data in Figure 2 show quite clearly that Equation \ref{mboltip}
constitutes an upper envelope to the distribution of the
observational points (with the exception of one cluster, namely NGC6352; in this case,
according to FPC83, the star considered to be
at the TRGB could also be a field star. The second more luminous
observed RGB star is $\approx$ 0.3 mag fainter), that are all
contained within 0.4-0.5 mag from Equation \ref{mboltip}.
This is exactly what expected on the basis of simple statistical
arguments (see Salaris \& Cassisi 1997), when
the evolutionary times in the upper part of the RGB and the
number of stars observed in each cluster are taken into account.
This means that the theoretical TRGB and ZAHB distance scales in GCs
are in agreement within the statistical uncertainties due to the small sample of
red giant stars observed.
\begin{figure}
\epsfxsize=6.5cm % fix the x-dimension and scales y-dim. to x-dim.
\hspace{3.9cm}\epsfbox{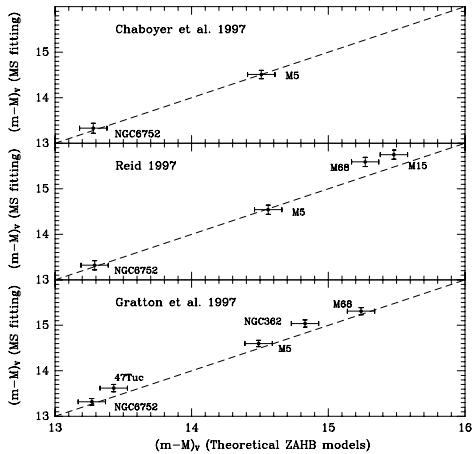} 
\caption[h]{Comparison between different GC distance moduli obtained by using
the MSF technique with HIPPARCOS subdwarfs, and our ZAHB distances (Equation \ref{rryale}).}
\end{figure}

Once assessed the consistency between TRGB and 
ZAHB distance scale for GC, we have
compared our GC ZAHB distances with distance moduli taken from the recent literature, 
derived from the Main Sequence Fitting (hereinafter MSF)
technique using subdwarfs with accurate HIPPARCOS parallaxes.
The sources of the GC MSF distances are Gratton
et al. (1997), Reid (1997) and Chaboyer et al. (1997).

In Figure 3 (panels a-c) we display the results of this comparison;
the error bars on the MSF distances are taken from the
quoted papers.
Due to the different procedures adopted by the various authors, 
the differences between the distance moduli obtained for the GCs in common
among these three investigations give us a rough estimate of the intrinsic error 
of the MSF technique.

It is worth noticing that on average there is a good agreement
between our ZAHB distance scale and the HIPPARCOS MSF distances.
For the most metal poor (and more distant) clusters displayed in the
figure, the Reid (1997) data seem to disagree systematically with our ZAHB
distances, but on the contrary the same M68 distance derived
by Gratton et al. (1997) nicely agrees with the ZAHB distance.
In particular, in the case of M5 and NGC6752 the distance moduli derived from the MSF 
by the three different groups are almost identical,
and the agreement with the ZAHB distance scale is almost perfect.
We can therefore finally conclude that for GC the HIPPARCOS, TRGB and
ZAHB distance scales are in agreement one with each other within
the present errors.

Once assessed the consistency between TRGB, ZAHB and HIPPARCOS
distances, 
the distance scale set by Equation \ref{rryale} can be also used for
reliably calibrating a relation providing [M/H] as a 
function of $(V-I)_{0,-3.5}$, by adopting a sample of GC for
which V-(V-I) C-M diagrams (as given by DA90)
and accurate spectroscopical [M/H] measurements (as listed in Salaris \&
Cassisi 1996) are available; this relation is needed for applying the
TRGB method to resolved galaxies.
We obtain:
\begin{equation}
[M/H]=-39.27 + 64.69\cdot[(V-I)_{0,-3.5}] - 36.35\cdot[(V-I)_{0,-3.5}]^2+ 6.84\cdot[(V-I)_{0,-3.5}]^3  
\label{vi35}
\end{equation}

\subsection{Resolved galaxies}

In the case of resolved galaxies, we can compare the TRGB distance scale
with the Cepheid and the RR Lyrae ones.
The observational database used in this comparison is the one in
Salaris \& Cassisi (1997), 
with the additional data for Sextans B taken from Sakai, Madore \&
Freedman (1997). 
In Table 2 we report the distance modulus determinations as obtained with the 
three different methods. 
The various columns provide the following data: (1) the name of the object;
(2) the reddening; (3) the observed I magnitude of the TRGB;
(4) the mean RGB metallicity, as obtained by adopting Equation \ref{vi35} and the
distance moduli in column 8;
(5) the intrinsic Cepheid
distance on the scale by Madore \& Freedman (1991), with the zero point
set by a LMC distance modulus of 18.50 mag and $\rm E(B-V)=0.10$;
(6) the true distance obtained by using the mean RR Lyrae V magnitude
(for details about the conversion from mean RR Lyrae brightness to the
corresponding ZAHB one see Cassisi \& Salaris 1997. When only the {\sl
g} magnitude of RR Lyrae stars is determined, it has been transformed
to V according to the relation by Kent 1985); 
(7) as in column (6) but for an average metallicity of the RR Lyrae population
[M/H]=-1.5 (see below); (8) the distance modulus obtained by applying
the TRGB method and making an average between the values obtained by using
the three different bolometric correction scales (see section 2.2).
The typical errors on the TRGB,
Cepheids and RR Lyrae distances for the sample of galaxies in Table 2 are 
on average of the order of 0.15 mag.

It is important to remember that the [M/H] values given in column 4 of Table 2 
are derived from
RGB stars, and correspond to an average metallicity of this stellar
population, that for the sample of galaxies in Table 2 shows generally a 
spread in [M/H] (this spread does not introduce a big error on the TRGB
distances, since the weak dependence of $M_{I}^{TRGB}$ on the metallicity).
In principle this average RGB metal content could be different from the RR
Lyrae one, especially for the highest and lowest values of [M/H] displayed
in Table 2,  
due to the low probability that metal-poor and 
metal-rich RGB stars evolve during their He central burning phase
through the RR Lyrae instability strip.
For roughly estimating the uncertainty due to the unknown original
metal content of the
RR Lyrae population, the distance moduli obtained assuming for the RR Lyrae stars 
an average metallicity equal to [M/H]=$-1.5$ - adopted as a reasonable estimate of the average
metallicity for the galactic halo RR Lyrae population - have been
reported in column 7 of Table 2 (with the unique exception of the LMC; in this
case we have a determination for the metallicity of the considered RR Lyrae).

When comparing TRGB distances determined with the three $BC_{I}$ 
scales presented in section 2.2, we obtain that the
average difference adopting
respectively the DA90 and Yale bolometric correction 
scales is $(m-M)_{TRGB,DA90}-(m-M)_{TRGB,Yale}$=-0.06$\pm$0.06 mag, while the
average difference when considering the DA90 and the K97 $BC_{I}$ is
$(m-M)_{TRGB,DA90}-(m-M)_{TRGB,K97}$=-0.08$\pm$0.06 mag. 
As already discussed in section 2.2, 
these differences can be considered as a rough estimate of the error on the TRGB
distances due to the uncertainty on the bolometric correction scale.
Moreover, the values of  $(m-M)_{TRGB,DA90}-(m-M)_{TRGB,Yale}$ and 
$(m-M)_{TRGB,DA90}-(m-M)_{TRGB,K97}$ are fully compatible with the
dispersion of the observational points around the empirical $BC_{I}$
scale by DA90 (see discussion in section 2.2).
In table 2 and in the following we have adopted, for
the TRGB distance of each galaxy, the value obtained by
averaging the three
distance moduli corresponding to the three different $BC_I$ scales.

A comparison between TRGB and RR Lyrae distances 
(when one neglects the very discrepant point corresponding
to NGC205; see also the discussion in LFM93 about this galaxy),
considering for the RR Lyrae the same mean
metallicity of the RGB stars, gives an average 
difference $(m-M)_{TRGB}-(m-M)_{RRLyrae}$=0.07$\pm$0.09 mag; when
considering (with the exception of the LMC) a 
metallicity [M/H]=-1.5, one obtains an average difference 
$(m-M)_{TRGB}-(m-M)_{RRLyrae}$=0.02$\pm$0.09 mag. 
One can therefore conclude that the RR Lyrae and TRGB distance scales
agree well, at the level of less than 0.10 mag, when considering our sample of
resolved galaxies.

In Figure 4, we have displayed the difference between the distance moduli obtained by adopting
the TRGB and the Cepheid distance scale. 
The average difference $(m-M)_{TRGB}-(m-M)_{Cepheids}$ between the two scales is equal to
0.12$\pm$0.06 mag, the TRGB distances being systematically larger,
in good agreement with the difference obtained considering only the LMC.
Since the good agreement between ZAHB and TRGB distances, and the
agreement between ZAHB and HIPPARCOS distance scales discussed in the
previous section, 
this systematic offset between the TRGB and Cepheid distance scales
supports, within the limits of the small sample of galaxy considered, 
the results by Feast \& Catchpole (1997), Gratton et
al. (1997), Reid (1997), that point to the direction of a higher LMC distance modulus 
(and higher zero point of the Cepheid distance scale) with
respect to the value of 18.50 mag adopted by Madore \& Freedman
(1991).

\vspace{0.2cm} 
\centerline{\bf Tab. 2 - Selected parameters for a sample of resolved galaxies.}
\begin{table}[h]
\hspace{1.0cm} %if you want to center your table act on this argument
\begin{tabular}{|lccc|cccc|}
\hline
   &  &  &  & \multicolumn{4}{|c|}{$(m-M)_0$}\\
\hline
Galaxy & $E(B-V)$ & $I_{Tip}$ & $[M/H]$ & ${Ceph}$ & 
${RR}$ & ${RR}^{-1.5}$ & ${Tip}$ \\
\hline
LMC     & 0.10  &14.60 &-1.0&18.50 &18.54 &      & 18.63\\
NGC6822 & 0.28  &20.05 &-1.7&23.62 &      &      & 23.67\\
NGC185  & 0.19  &20.30 &-1.0&      &24.06 &24.15 & 24.14\\
NGC147  & 0.17  &20.40 &-0.9&      &24.17 &24.28 & 24.28\\
IC1613  & 0.02  &20.25 &-1.2&24.42 &24.41 &24.46 & 24.45\\
M31     & 0.08  &20.55 &-0.9&24.44 &24.56 &24.67 & 24.60\\
M33     & 0.10  &20.95 &-2.0&24.63 &24.85 &24.78 & 24.90\\
WLM     & 0.02  &20.85 &-1.5&24.92 &      &      & 25.03\\
NGC205  & 0.035 &20.45 &-0.9&      &24.90 &25.01 & 24.59\\
Sex A   & 0.075 &21.79 &-1.9&25.85 &      &      & 25.92\\
Sex B   & 0.015 &21.60 &-1.6&25.69 &      &      & 25.79\\
NGC3109 & 0.04  &21.55 &-1.5&25.50 &      &      & 25.68\\
\hline
\end{tabular}
%%%%%%%%%%%%%%%%%%%%%%%%%%%%%%%%%%%%%%%%%%%%%%%%%%%%%%%%%%%%%%%%%%%%%%%%%
\end{table}
%%%%%%%%%%%%%%%%%%%%%%%%%%%%%%%%%%%%%%%%%%%%%%%%%%%%%%%%%%%%%%%%%%%%%%%%%%
%
%
%
\begin{figure}
\epsfxsize=6cm % fix the x-dimension and scales y-dim. to x-dim.
\hspace{4.5cm}\epsfbox{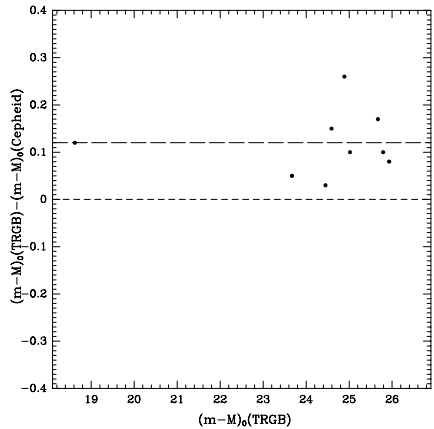} 
\caption[h]{Comparison between different distances for the selected
sample of resolved galaxies, obtained by using the TRGB 
and the Cepheid distance scales. The long dashed line corresponds to the average 
difference between the TRGB and Cepheid distance moduli.}
\end{figure}

\section{The Leo I group and Coma cluster distances, and the value of $H_{0}$.}

The Leo I group 
is a relatively nearby group of galaxies, compact, with a
line-of-sight depth 
estimated to be $\approx$2\% compared to its distance (Tanvir et al. 1995).
Very recently, Sakai et al. (1997 - hereinafter SA97) 
detected the TRGB in NGC3379 (one of the dominant galaxies in Leo I), by
means of $HST$ WFPC2 observations. They placed the observed TRGB at
I=26.32$\pm$0.05 mag, assumed $A_{I}$=0.02 mag, and adopted a 
metallicity [M/H]=-0.68$\pm$0.40 (see SA97 for more details).

By using the quoted values 
(and the associated errors) for extinction, metallicity and TRGB
location, we derive a TRGB distance modulus 
$(m-M)_{0,3379}$=30.48$\pm$0.12 mag 
(other sources of errors included in the error budget are
the uncertainty on the WFPC2 photometric zero point as given by SA97, the
uncertainty on the theoretical calibration of the TRGB, estimated to be
of $\pm$0.05 mag on the base of the comparison in Figure 1, and the
variation of the TRGB brightness due to a variation $\Delta$Y=$\pm$0.03 in the
intial He content of the theoretical models, that is $\approx \pm$0.03
mag). 
This distance modulus corresponds to a linear distance
$d_{3379}$=12.5$\pm$0.7 Mpc, and it is $\approx$ 8\% higher than the value
derived by SA97, using the DA90 calibration of the TRGB distance scale.

Once fixed the absolute distance to Leo I, we can obtain the distance to
the Coma cluster using the distance ratio Coma-Leo I as determined by means of 
secondary distance indicators.

According to the recent analysis by Colless \& Dunn (1996) the Coma
cluster consists of two components; the main cluster 
centered around NGC4874 and NGC4889, with a mean heliocentric
recession velocity $cz$=6853 km $s^{-1}$, and a subgroup around NGC4839
characterized by a mean value of $cz$=7339 km $s^{-1}$. 
The relative distance between galaxies in the main component of Coma and
Leo I has been recently redetermined by Gregg
(1997) by means of the diameter - velocity dispersion method, and results
to be $d_{Coma}/d_{Leo I}$=8.84$\pm$0.23. The same relative distance  is also obtained 
when using the diameter - velocity dispersion data by Faber et al (1989) for 2 
ellipticals in Leo I and 27 ellipticals in the main component of the Coma
cluster.
With the TRGB Leo I distance modulus previously derived, this
Coma-Leo I distance ratio
provides a distance to the main compo\-nent of the Coma cluster
$d_{Coma}$=111$\pm$9 Mpc, and
$(m-M)_{0,Coma}$=35.23$_{-0.19}^{+0.17}$ mag (accounting in the error
budget also for an uncertainty by $\pm$0.04 mag due to the
r.m.s. depth of the Leo I group as given by Tanvir et al 1995).
Once the distance to Coma is known, the value of $H_{0}$ is derived using 
the heliocentric recession velocity of the main cluster component 
($cz$=6853 Km $s^{-1}$) 
transformed to the centroid of the Local Group,
and corrected for the motion of the Local Group relative to the
cosmic background radiation in the direction of Coma (272 Km $s^{-1}$
according to Staveley-Smith \& Davies 1989, to which we attach an
error by $\pm$100 Km $s^{-1}$).
Morever, we corrected for the peculiar motion ($V_{p}$) of the cluster 
as estimated by Han \& Mould (1992), 
$V_{p}$=+66$\pm$ 428 Km $s^{-1}$ (we have taken the median
value of their three solutions for $V_{p}$).
We finally obtain a cosmic
recession velocity $cz$=7068$\pm$440 Km $s^{-1}$, and
$H_{0}$=64$_{-9}^{+10}$ Km $s^{-1} Mpc^{-1}$.
\acknowledgements
The work of one of us (M.S.) was carried out as part of the TMR
programme (Marie Curie Research Training Grants) financed by the EC.

\end{document}